\documentclass[twocolumn, switch]{article} 

\usepackage{preprint}

\usepackage{amsmath, amsthm, amssymb, amsfonts}

\usepackage[numbers,square]{natbib}
\usepackage[utf8]{inputenc}	
\usepackage[T1]{fontenc}	
\usepackage{xcolor}		
\usepackage[colorlinks = true,
            linkcolor = purple,
            urlcolor  = blue,
            citecolor = cyan,
            anchorcolor = black]{hyperref}	
\usepackage{booktabs} 		
\usepackage{nicefrac}		
\usepackage{microtype}		
\usepackage{lineno}		
\usepackage{float}			
\usepackage{multirow}
\usepackage{lipsum}		

\usepackage{newfloat}
\DeclareFloatingEnvironment[name={Supplementary Figure}]{suppfigure}
\usepackage{sidecap}
\sidecaptionvpos{figure}{c}

\usepackage{titlesec}
\titlespacing\section{0pt}{12pt plus 3pt minus 3pt}{1pt plus 1pt minus 1pt}
\titlespacing\subsection{0pt}{10pt plus 3pt minus 3pt}{1pt plus 1pt minus 1pt}
\titlespacing\subsubsection{0pt}{8pt plus 3pt minus 3pt}{1pt plus 1pt minus 1pt}

\usepackage{tikz,xcolor,hyperref}

\definecolor{lime}{HTML}{A6CE39}
\DeclareRobustCommand{\orcidicon}{
	\begin{tikzpicture}
	\draw[lime, fill=lime] (0,0)
	circle [radius=0.16]
	node[white] {{\fontfamily{qag}\selectfont \tiny ID}};
	\draw[white, fill=white] (-0.0625,0.095)
	circle [radius=0.007];
	\end{tikzpicture}
	\hspace{-2mm}
}
\foreach \x in {A, ..., Z}{\expandafter\xdef\csname orcid\x\endcsname{\noexpand\href{https://orcid.org/\csname orcidauthor\x\endcsname}
			{\noexpand\orcidicon}}
}

\title{VPU-EM: An Event-based Modeling Framework to Evaluate NPU Performance and Power Efficiency at Scale}

\usepackage{xwatermark}

\usepackage{authblk}

\author[1]{Charles Qi}
\author[1]{Yi Wang}
\author[1]{Hui Wang}
\author[1]{Yang Lu}
\author[1]{Shiva Shankar Subramanian}
\author[1]{\authorcr Finola Cahill}
\author[1]{Conall Tuohy}
\author[1]{Victor Li}
\author[1]{Xu Qian}
\author[1]{Darren Crews}
\author[1]{Ling Wang}
\author[1]{Shivaji Roy}
\author[1]{\authorcr Andrea Deidda}
\author[1]{Martin Power}
\author[1]{Niall Hanrahan}
\author[1]{Rick Richmond}
\author[1]{\authorcr Umer Cheema}
\author[1]{Arnab Raha}
\author[1]{Alessandro Palla}
\author[1]{Gary Baugh}
\author[1]{Deepak Mathaikutty}

\affil[1]{Intel Corporation}

\usepackage{graphicx}

\usepackage{url}

\begin{document}

\twocolumn[ 
  \begin{@twocolumnfalse} 

    \maketitle

    \begin{abstract}
      State-of-art NPUs are typically architected as a self-contained sub-system with multiple heterogeneous hardware computing modules, and a dataflow-driven
      programming model. There lacks well-established methodology and tools in the industry to evaluate and compare the performance of NPUs from different architectures.
      We present an event-based performance modeling framework, VPU-EM, targeting scalable performance evaluation of modern
      NPUs across diversified AI workloads. The framework adopts high-level event-based system-simulation methodology to abstract away design details for speed, while
      maintaining hardware pipelining, concurrency and interaction with software task scheduling. It is natively developed in Python and built
      to interface directly with AI frameworks such as Tensorflow, PyTorch, ONNX and OpenVINO, linking various in-house NPU graph compilers to achieve optimized full
      model performance. Furthermore, VPU-EM also provides the capability to model power characteristics of NPU in Power-EM mode to enable joint performance/power analysis.
      Using VPU-EM, we conduct performance/power analysis of models from representative neural network architecture. We demonstrate that even though this framework is developed for
      Intel VPU, an Intel in-house NPU IP technology, the methodology can be generalized for analysis of modern NPUs.
    \end{abstract}
    \vspace{0.35cm}

  \end{@twocolumnfalse} 
] 



\section{Introduction}

With the explosion of AI, high-performance NPUs aimed at accelerating AI computation with power efficiency have emerged as an alternative solution
to CPUs and GPUs. CPUs and GPUs are built for non-AI or general purpose computing \cite {nickolls2009graphics}. AI solutions based on CPU or GPU exploit instruction level parallism for
speedy computation while maintaining flexible instruction set architecture(ISA). ISA simulators and profilers for CPUs and GPUs are well-developed
by academia and industry to evaluate the performance of CPU and GPU at architecture level \cite{bakhoda2009analyzing, binkert2011gem5} and \cite{ubal2012multi2sim}.

NPUs focus on maximizing computation density and power efficiency in AI computing. In order to achieve this, state-of-art NPUs are typically architected
as a self-contained sub-system with heterogeneous high-density hardware computing modules and embedded buffer memories
\cite{chen2016eyeriss,parashar2017scnn,talpes2020compute,liu2020systolic,8693202} and \cite{jouppi2021ten}.
A dataflow programming model is often deployed, which enables a neural network model to be compiled and optimized via neural network (NN) graph compilers \cite{li2020deep}
and then mapped to the computing modules of the NPU. Due to the complexity of NPU architecture and its programming model, it is a challenging task to conduct
architecture performance studies of NPU. To our knowledge, there lacks well-established methodology and tools to thoroughly evaluate NPU performance/power efficiency
prior to silicon commitment.

In this paper, we present VPU-EM framework, an event-driven modeling methodology for analyzing architecture trade-off and projecting performance/power efficiency of NPUs.
We start by analyzing the representative architecture characteristics of NPUs, using Intel Vision Processing Unit (VPU) IP as an example.
It is demonstrated that state-of-art NPUs like Intel VPU are often built as a self-contained sub-system with multiple computing units, embedded memory, a built-in DMA and
a management processor. The computing units include hardware-oriented high-density MAC arrays as well as programmable components such as VLIW DSPs.
Based on these characteristics, we examine the limitations of the traditional ISA simulator based approach for evaluating NPU performance. We therefore target the VPU-EM framework
for NPU architecture specifically, with several modeling objectives outlined to make NPU performance projection more feasible under this framework.
With regard to power efficiency optimization of NPUs, our paper emphasizes the importance to conduct joint performance/power analysis using real AI workloads. We further
demonstrate how this can be accomplished with the Power-EM simulation mode of VPU-EM.

We then present the detailed design of the VPU-EM framework to achieve these objectives, including:
\begin{itemize}
  \item Event-driven methodology and coding language
  \item Modeling hardware components
  \item Modeling processing flow
  \item Characterize framework accuracy
\end{itemize}

Utilizing the VPU-EM framework, we conduct several performance analyses on Intel VPU with a wide range of design parameters
for computation scaling, frequency scaling and memory BW scaling. We demonstrate that the VPU-EM framework is highly scalable and flexible.
It enables systematic performance analyses of generations of Intel VPU architecture and provides sufficient accuracy when correlated with the actual
design implementation.

In order to establish key advantages on power efficiency in the early architecture definition phase of the VPU, we recognize the importance for the VPU-EM framework
to provide the capability to conduct joint performance/power analysis seamlessly.
We present a power simulation mode for VPU-EM, Power-EM, to analyze the detailed power characteristics of VPU hardware modules simultaneous with performance analysis
under the same AI workloads. This allows key power efficiency metrics such as inference/Watts, eTOPS/Watts, etc. to be established during the architecture definition
phase. It also provides guidance to define dynamic power management features, such as active power state management, DVFS, etc.

The remainder of the paper is organized as follows. Section \ref{sec:background} provides the background that motivates our work.
Section \ref{sec:design} provides detailed descriptions of the VPU-EM framework. Section \ref{sec:performance} demonstrates how to use the framework to conduct various
performance analyses of NPU architecture.
Section \ref{sec:power} provides description of Power-EM simulation mode and highlights the capability of VPU-EM to conduct joint performance/power studies of NPU.
Section \ref{sec:conclusion and future} draws conclusions for this paper and outlines some future work.

\section{Background}
\label{sec:background}

NPUs deploy unique hardware architecture features with significant performance implications. These unique features cannot be sufficiently evaluated using
a traditional ISA simulator based approach found in CPU or GPU studies. We analyze the uniqueness of NPU architecture using Intel VPU and explain the shortcomings
of ISA simulator in order to establish the modeling objectives of VPU-EM.

\subsection{Uniqueness of NPU Architecture}
\label{subsec:architecture}
NPUs first emerged in embedded devices as power-efficient AI inference accelerators to handle the computational demand of convolutional neural network(CNN)
\cite{bianco2018benchmark}. Initially NPU implementation focused on the design of a high-density MAC array, constructed as a 2-dimensional GEMM or a 3-dimensional systolic
array \cite{chen2016eyeriss,parashar2017scnn,talpes2020compute,liu2020systolic,8693202} and \cite{jouppi2021ten}, because over 90\% of the CNN computational workload is generated
by the convolution layers.

As the CNN architecture evolves with deeper layers and more complex computation functions, the MAC array architecture of NPUs has evolved with greater flexibility to maintain
high utilization and to scale up performance, including features such as:
\begin{itemize}
  \item Multi-level array partition
  \item Adjustable array dimensions
  \item Tight-coupling with high BW memories
  \item Flexible resource allocation and dataflow
  \item Mixed precision and sparsity acceleration support
  \item Fused operation support
\end{itemize}

NPUs are also adapted to support NN architectures other than CNN. RNN/LSTM architecture \cite {sutskever2014sequence} has been widely deployed in AI-enabled audio/speech applications.
Most recently transformer architecture \cite {han2022survey} has emerged as a promising architecture for AI. New generations of NPUs enables broader NN support
by integrating flexible computing elements such as VLIW DSP, special math function hardware modules,
or memory-to-memory data transformation modules \cite {NVDLA-doc}, increasing heterogeneity of NPU architecture even further.

AI processing is memory bandwidth (BW) and capacity intensive. Memory technology does not scale with the semiconductor process to keep up with AI computation demands. NPUs strive to maintain
high utilization of hardware computing resource by deploying highly customized multi-level memory hierarchies. Furthermore, NPUs often deploy custom tensor-aware DMAs with software managed
memory accesses or data compression techniques \cite {rhu2018compressing} to minimize memory BW demand and access latency.

\begin{figure}[H]
  \centering
  \includegraphics[scale=0.38]{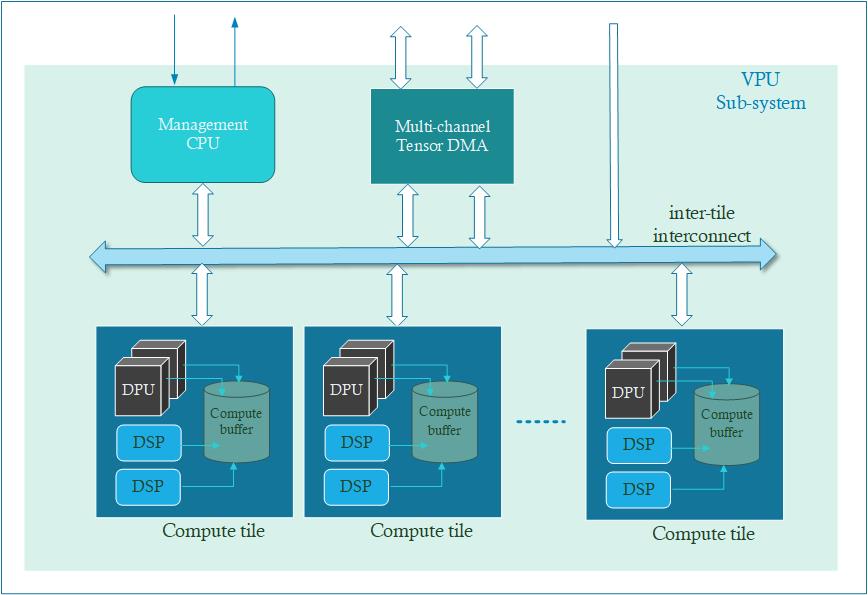}
  \caption{NPU Architecture - Intel VPU}
  \label{fig:fig1}
\end{figure}

The above characteristics of NPU are embodied in the architecture of Intel VPU IP, as shown in Figure \ref{fig:fig1}. Without losing generality, the VPU architecture contains
multiple computing tiles for scaling, connected via an inter-tile interconnect. Each compute tile contains multiple MAC arrays (DPUs) and multiple VLIW DSP processors,
sharing a high-bandwidth local RAM. The VPU contains a management processor to interface with the host driver for processing request scheduling and task management. The VPU
also contains a multichannel, tensor-aware DMA to access host DDR memory with the support of data compression/decompression to reduce memory BW. The VPU exposes a host programming
interface to map the internal RAMs and registers to host address space and enumerates itself as a virtual PCIE accelerator device.

\subsection{Related Work}
\label{subsec:related}

ISA simulators have been widely used to analyze the performance of CPUs and GPUs \cite{bakhoda2009analyzing, binkert2011gem5} and \cite{ubal2012multi2sim}.  Most recently
GPU simulators like GPGPU-sim are enhanced with modeling of AI specific processing modules, e.g. NVIDIA Tensor core or direct support of AI frameworks e.g. PyTorch
\cite{aamir2018modeling, lew2019analyzing}. The main function of an ISA simulator is to simulate the execution of software codes that have been compiled into instructions,
using a processor model. ISA simulators focus on kernel level optimization and the execution and interaction of multiple instruction streams on
a multithreading machine. It is difficult to represent NPU hardware acceleration functions controlled by atomic hardware state machines as sequential instructions simulated
via ISA simulators. Another disadvantage of ISA simulators
is that they typically simulate both functional accuracy and cycle count performance simultaneously, making the simulation speed extremely slow for full model inference.
Furthermore, ISA simulators are often found to be inaccurate in GPU performance projections due to insufficient modeling of GPU memory hierarchy \cite{jain2018quantitative}.

\begin{figure}[H]
  \centering
  \includegraphics[scale=0.25]{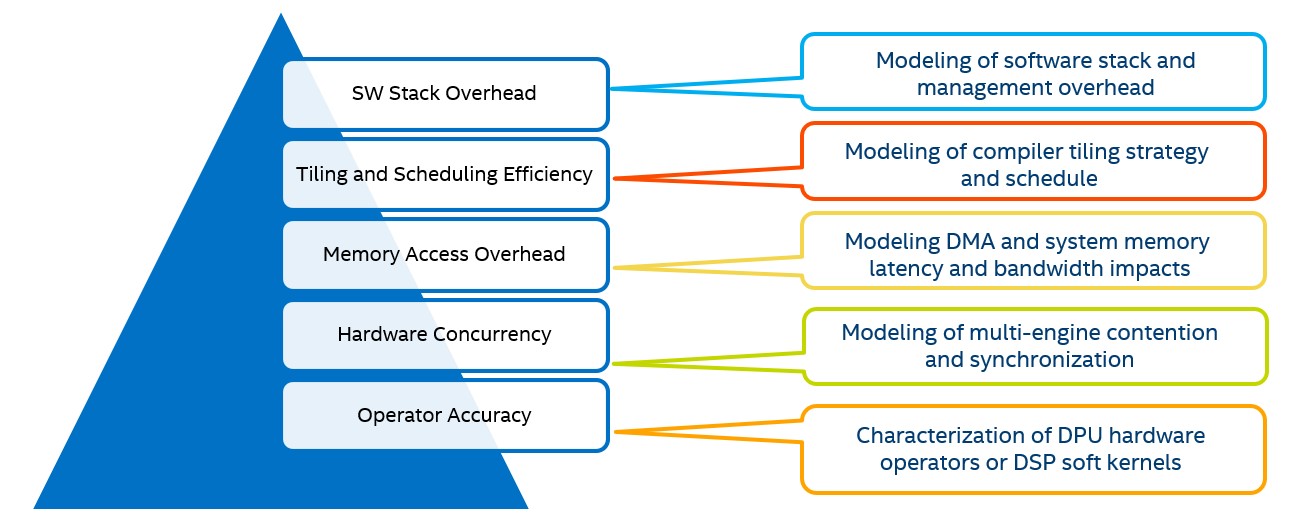}
  \caption{Key Modeling Factors for Accuracy}
  \label{fig:fig2}
\end{figure}

With the emergence of NPU, simulators have been developed to focus on the performance study of MAC arrays \cite{samajdar2018scale, wu2022sparseloop}. Such simulators provide great insights to the architecture
trade-offs between MAC array design parameters. But it is insufficient to assess full model or use case level performance of the NPUs. As illustrated by the VPU architecture,
several factors (Figure \ref{fig:fig2}) contribute to the accuracy of performance projections which must be taken into consideration in the performance model.


\subsection{Modeling Objectives}
\label{subsec:objectives}

Recognizing the challenges we face to conduct thorough and systematic performance analysis of Intel VPU architecture, we set several development objectives for the VPU-EM
framework.

\paragraph{Workload Diversity} VPU-EM SHALL accept a wide range of AI models developed from multiple machine learning (ML) frameworks.

\paragraph{Parameter Scaling} VPU-EM SHALL provide flexibility to permute a large set of design parameters for architecture trade-off analysis.

\paragraph{Overhead Analysis} VPU-EM SHALL model and quantify the impact of system memory latency and BW.

\paragraph{Insightful Data Analytics} VPU-EM SHALL provide detailed performance trace and report capability.

\paragraph{Simulation Speed} VPU-EM SHALL simulate typical AI models (e.g. Resnet50 224x224) within several minutes.

\paragraph{Operator Accuracy} VPU-EM SHALL provide NPU specific operator mapping and modeling within 5\% to 10\% accuracy.

\paragraph{Power Analysis} VPU-EM SHALL provide capability to conduct joint performance/power analysis.


\section{Framework Design}
\label{sec:design}

In this section, we provide detailed design of the VPU-EM framework in order to achieve the outlined objectives.

\subsection{Event-driven Methodology}

The most efficiency way for VPU to accelerate AI computing is to offload a full model inference as a single processing request.
So the top priority of the VPU-EM framework is to provide direct support for a wide range of representative AI models and speedy simulation to explore large
design parameter space using these models.

Traditionally performance modeling is done using C++ or SystemC as cycle-approximate or cycle-accurate models. The development cycle for such a model
is very long, and the simulation speed is very slow. A full model inference simulation on a cycle-oriented ISA simulator may take
several hours or 1-2 days to complete (still considered faster than RTL simulation, which may take days or weeks).
Furthermore, it is difficult to integrate the C++/SC models directly with a ML framework like Tensorflow or PyTorch.
Studies based on these models are often limited to operator level with manually created stimulus.

\subsubsection{Event-based Simulation}
\label{subsubsec:event sim}

To speed up full model simulation, VPU-EM borrows the event-driven simulation concept often seen in large system simulations. It simulates only time-critical events
which capture the performance characteristics of the VPU. Events are defined in two levels in VPU-EM with the second level being equivalent
to SC TLM transactions,

\begin{itemize}
  \item Task level events to capture interaction of task scheduler and processing engines
  \item Sub-task level events to simulate hardware pipeline and latency within an engine
\end{itemize}

\subsubsection{Python Language}
\label{subsubsec:Python}
Furthermore, we made a conscious decision to select Python as the primary coding language for VPU-EM. This decision is driven by several factors:
\begin{itemize}
  \item Many ML frameworks provide Python binding and rich libraries to process AI models
  \item Python is easy to interface in-house graph compilers
  \item Python reduces the time to develop and extend the infrastructure
  \item Python is convenient to debug and conduct data analytics
\end{itemize}

\subsubsection{SimPy Library}
\label{subsubsec:simpy}

In order to model hardware events and concurrent execution of hardware models, we need to develop a modeling infrastructure to represent
hardware processes, handshake signals, FIFOs and queues, as well as track and advance simulation cycles. In SystemC, this is provided
by multithreading and modeling class libraries of the language (e.g. sc\_thread, sc\_time, sc\_fifo).

In order to build the modeling infrastructure in Python, we select a third party system level simulation library called SimPy \cite{simpy-ppt, simpy-doc}.
SimPy defines a simulation environment to track events and simulate time advancement. It utilizes Python generator capability to model
the interaction of multiple concurrent processes. SimPy also provides various types of shared resources.

In VPU-EM, predefined SimPy classes are leveraged to rapidly develop hardware-oriented modeling components,
\begin{itemize}
  \item SimPy environment class is leveraged to construct VPU-EM testbench and launch simulation
  \item SimPy store class is leveraged to construct hardware FIFOs and queues
  \item SimPy container class is leveraged to construct shared memory
  \item SimPy process class is leveraged to construct concurrent hardware modules and state machines
  \item SimPy event class is leveraged to create hardware handshake signals such as interrupt
\end{itemize}


\subsection{Modeling Hardware Components}
\label{subsec:hardware}

The modeling methodology of key hardware components is described below.

\begin{figure}[H]
  \centering
  \includegraphics[scale=0.36]{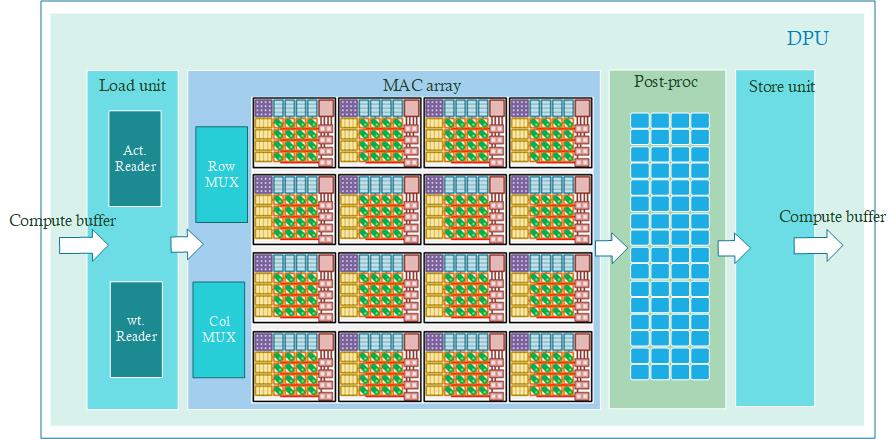}
  \caption{DPU Pipeline}
  \label{fig:fig3}
\end{figure}

\paragraph{DPU}
In VPU, operators such as convolution, depthwise and transpose convolution, and matrix multiplication are supported by DPU MAC arrays.
The MAC array is organized into a two-dimensional array of PE cells. Each PE cell supports dot-product computation in channel dimension for up to 16
MACs. The MAC array supports flexible configuration of activation or weight reuse with local context buffers. The load and store units support high BW
activation/weight read access and result writeback to compute buffer respectively. DPU also supports fused activation, elementwise add and
batch normalization in a separate post-processing stage (Figure \ref{fig:fig3}).

The DPU is modeled as a 4-stage pipeline, load, MAC array, post-processing and store. In the DPU model, we design the unit of processing as a data block flowing through the pipeline,
to reflect compute-bound vs. memory-bound performance characteristics. Rather than modeling cycle-level hardware control, we focus the modeling effort on representing
the unique DPU architecture for maximizing data reuse in the computation through flexible partitioning of the activation and weight tensors called stencils.
For a given opcode and its associated input/output tensors, the size of the data block is dynamically decided to be a sub-partition
of the tensor sizes that are multiples of the selected stencil configuration. The full operator is modeled as multidimensional outer loops on top of the data block.
The operator-specific data block partition and the pipelined model allows the DPU hardware performance to be projected with sufficient accuracy.

\begin{figure}[H]
  \centering
  \includegraphics[scale=0.50]{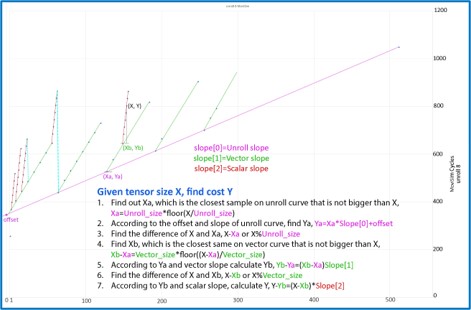}
  \caption{DSP Kernel Characterization}
  \label{fig:fig4}
\end{figure}

\paragraph{DSP}
The VPU contains VLIW DSPs to process special operators. In VPU-EM, the DSP is modeled as a three-stage pipeline.
The unit of processing is a data block configurable as multiple SIMD vectors. In order to achieve accuracy for VLIW architecture, we utilize MoviSim ISA simulator
to characterize DSP kernels offline into parameterized lookup tables. We build template-based kernels with hand-tuned vectorization and loop unrolling.
For example, it is observed that elementwise nonlinear functions, (tanh, sigmoid, hswish etc.) can be represented by one offset and three linear curves.
The offset represents the preamble to prepare and initialize the kernel. The linear curves represent multiples of loop unrolling block, SIMD vector
and scalar respectively. Using the characterization table (Figure \ref{fig:fig4}) and the three-stage pipeline model, DSP kernel performance can
be simulated with high accuracy.

\paragraph{Compute Buffer Memory}
The local RAM embedded in each VPU compute tile, also known as the compute buffer (CB) is a multi-port high bandwidth(BW) memory which serves as the main daa memory
for DPU and DSP. Its performance is critical to the overall performance of VPU. In VPU-EM, the CB is modeled as a multi-port memory with
configurable BW and latency parameters matching the actual implementation. The CB model connects to the load/store pipeline stages
of the DPUs and DSPs. It also provides additional ports for DMA and inter-tile communication.

\paragraph{DDR Memory}
The DDR memory model is built using the same base class memory model. However, it also models performance-critical DDR functionalities based on selected DDR
standards. These include DDR timing parameters, burst length, bank configuration, page size, refresh modes etc. Furthermore, the DDR model supports
translating linear addresses into DDR device addresses with bank interleaving and page policy management \cite {blackmore2013quantitative}.

\paragraph{DMA}
The VPU DMA is a multichannel tensor-aware DMA. It supports complex memory access patterns representing multidimensional tensor storage layout in memory.
The DMA can perform memory-to-memory data transfers between DDR-DDR, DDR-CB or CB-CB based on the descriptors prescribed by the NN compiler. It can also
perform additional inline data processing steps during data transfer, including compression/decompression, transpose and permutation. For efficient data transfer,
the DMA has broadcast capabilities to distribute the same data to multiple compute tiles. VPU-EM provides a detailed DMA model following the actual RTL design.
However, the DMA model abstracts away the cycle-level details and focuses on multi-agent data transfer characteristics.
It models how a DMA descriptor is split into pipelined data transfer requests. For each request, it projects latency and BW data. The data is aggregated to provide the final
result of a DMA task.

\paragraph{Interconnect}
The inter-tile interconnect of VPU is modeled using a parameterized generic NOC model consisting of multiple slave and master ports, and a centralized
router module to forward requests and responses between the slave and the master ports. The router model supports address-based or ID-based unicast or multicast
routing. It also supports commonly used arbitration schemes. Latency and BW parameters are configurable for the model to reflect the performance impact of the interconnect.
In VPU-EM testbench, the same NOC model is also used to construct the SOC-level interconnect between VPU and DDR models.

\subsection{Modeling Processing Flow}
\label{subsec:processing}

The modeling methodology of the processing flow is described below.

\paragraph{Parameter Configuration}

Configuration parameters are defined hierarchically using yaml files and imported into configuration class objects. They are used by
hardware module class constructors or during simulation to steer the performance analysis runs.
Configurations parameters capture what can be adjusted through hardware registers in a given implementation as well as design space parameters
to make trade-off analysis.

\paragraph{Operators and Tasks}

Operators and tasks are class objects derived from base classes extensible through a factory mechanism of Python.
In VPU-EM, operators are defined following OpenVINO IR opset. But they can be flexibly mapped to different processing engines. VPU-EM
defines both computing and DMA tasks. A computing task may contain a partial operator from tiling or multiple
operators fused together. A DMA task contains a complex DMA request defined by one or more DMA descriptors.

\paragraph{Scheduling and Synchronization}
The unit of scheduling in VPU-EM is a task. A centralized scheduler connects to different hardware engines via task FIFOs. The scheduler
parses an AI model into a task list and enqueues the tasks into the FIFOs when there is room. Tasks are processed asynchronously by the engines. The scheduler tracks
the completion of the tasks in separate threads.

Data or resource dependencies of the tasks are resolved through a barrier mechanism. Logical barriers are inserted by the NN compiler into an
AI models. VPU-EM contains a barrier scoreboard model to track the state of each barrier.
Barriers contain semaphore counters and can generate globally observable events. Engines form produce-consumer relationship to synchronize
task processing atomically based on barrier state.




\subsection{Modeling Accuracy Characterization}
\label{subsec:accuracy}
We characterize the accuracy of the VPU-EM performance projection results using RTL simulation data from three AI models in design validation sign-off.
We also compare these results against a neural network cost model, VPUNN, trained with ground truth data from FPGA measurements.
We select four configurations for each model, original, with compression, with sparsity acceleration, with both sparsity acceleration and
compression. The RTL simulation environment assumes a more ideal system memory behavior. The FPGA data for the cost model is collected at operator level.
The comparison result shows (Table \ref{tab:tab1}) that VPU-EM tracks RTL and VPUNN results with 5\% accuracy for all original models.
For sparse and sparse-compressed models, further enhancement of VPU-EM is required but VPU-EM results fall between RTL and VPUNN results.


\begin{table}[H]
  \caption{Accuracy Characterization}
  \centering
  \begin{tabular}{lccc}
    \toprule
    \textbf{Model Name} & \textbf{VPUNN}   & \textbf{VPU-EM}  & \textbf{VPU-EM}    \\
    \textbf{}           & \textbf{vs. RTL} & \textbf{vs. RTL} & \textbf{vs. VPUNN} \\
    \midrule
    {MobileNet v2	 }     & 2\%              & 3\%              & 1\%                \\
    {MobileNet v2\_C}   & 2\%              & 3\%              & 1\%                \\
    {MobileNet v2\_S}   & 2\%              & 7\%              & 5\%                \\
    {MobileNet v2\_SC}  & 2\%              & 7\%              & 4\%                \\
    {ResNet50}          & 4\%              & 0\%              & -4\%               \\
    {ResNet50\_C}       & 3\%              & -1\%             & -4\%               \\
    {ResNet50\_S}       & -31\%            & -15\%            & 20\%               \\
    {ResNet50\_SC}      & -32\%            & -15\%            & 20\%               \\
    {Tiny YOLO v2}      & -5\%             & -10\%            & -5\%               \\
    {Tiny YOLO v2\_C}   & -6\%             & -11\%            & -5\%               \\
    {Tiny YOLO v2\_S}   & -42\%            & -19\%            & 29\%               \\
    {Tiny YOLO v2\_SC}  & -43\%            & -19\%            & 30\%               \\
    \bottomrule
  \end{tabular}
  \label{tab:tab1}
\end{table}

\section{Performance Analysis}
\label{sec:performance}

In this section, we present example analysis to demonstrate the capabilities of the VPU-EM framework. The data presented does not represent the actual KPI of VPU. The configurations
are purposely skewed to highlight the impacts on performance from various architecture decisions quantifiable via the framework.

\subsection{Computation Scaling}
\label{subsec:computation}
\begin{figure}[H]
  \centering
  \includegraphics[scale=0.50]{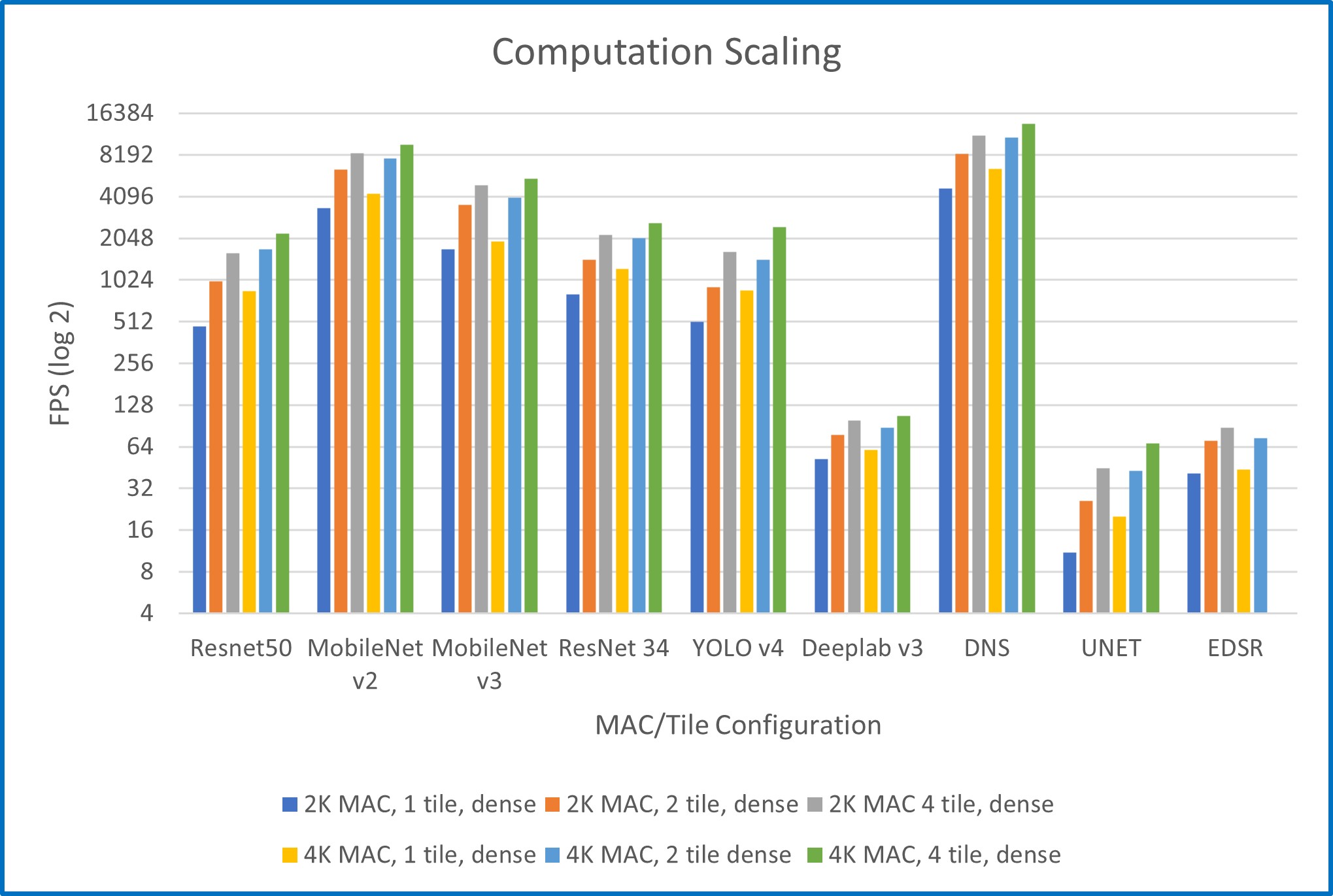}
  \caption{Computation Scaling Analysis}
  \label{fig:fig6}
\end{figure}

In this analysis, we project the performance of representative AI models for multiple MAC array and compute tile configurations. As shown in Figure \ref{fig:fig6},
for a single model, we can achieve an average 1.9x of scaling from one tile to two tiles. However, the scaling factor
drops to about 1.47x going from two tiles to four tiles. It is also shown that increasing the array from 2K to 4K MACs alone only improves performance by about 25\%-45\%,
suggesting lower array utilization and insufficient scaling of other VPU resources.

\subsection{Frequency Scaling}
\label{subsec:frequency}
\begin{figure}[H]
  \centering
  \includegraphics[scale=0.65]{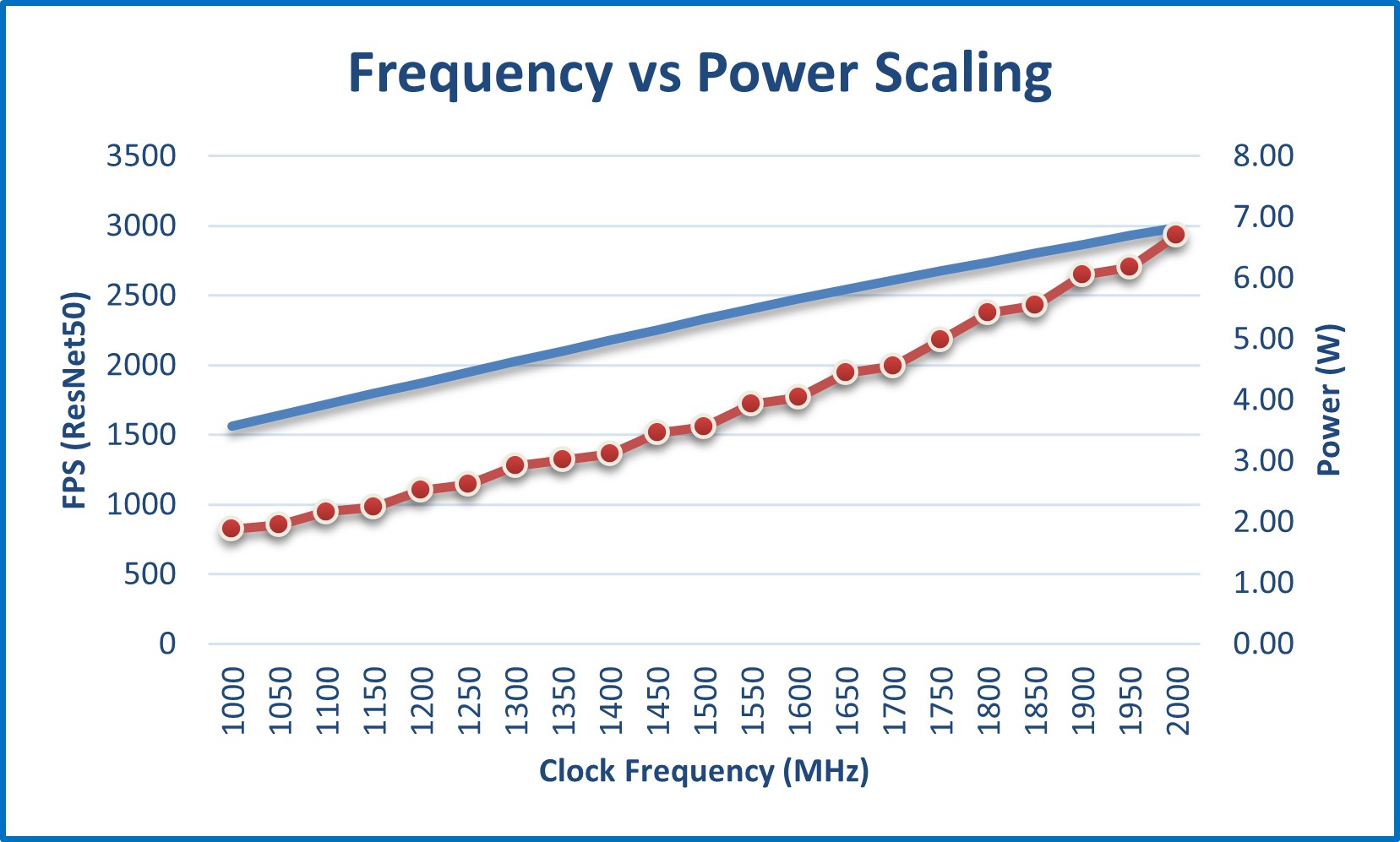}
  \caption{Frequency Scaling Analysis}
  \label{fig:fig7}
\end{figure}

In this analysis, we analyze the performance of VPU vs. clock frequency scaling and correlate with the power consumption of VPU. It is shown in Figure \ref{fig:fig7}
that the performance of VPU scales linearly with increasing frequency, while the power consumption increases at a faster rate due to increasing voltage. So VPU
is more power efficient when operating at a lower frequency.

\subsection{Memory BW Scaling}
\label{subsec:memory BW}
\begin{figure}[H]
  \centering
  \includegraphics[scale=0.50]{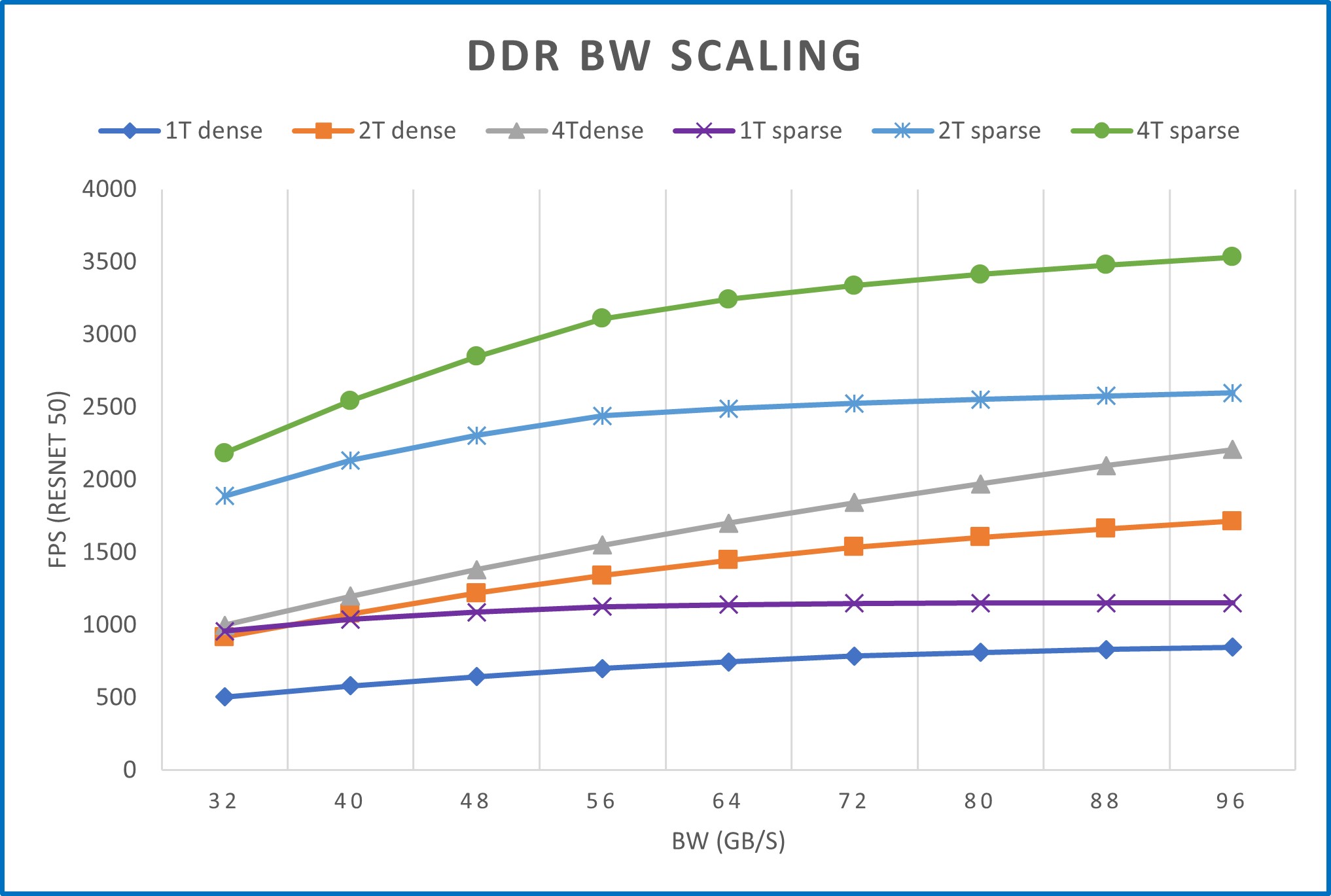}
  \caption{Memory BW Scaling Analysis}
  \label{fig:fig8}
\end{figure}

In this analysis, we analyze the performance of VPU vs. DDR memory BW scaling. It is shown in Figure \ref{fig:fig8} that DDR BW has significant impact on VPU Performance
for dense models, and for design configurations with limited compute buffer.

\section{Power Analysis}
\label{sec:power}
State-of-art NPUs strive for power-efficiency as their key differentiation. Therefore, it is important to conduct performance and power analysis jointly in early
NPU architecture development. Previous academic research works in \cite{854380, 5375438, 6853196} and \cite{leng2013gpuwattch} provide good guidance on the methodology
to conduct architecture level power analysis. However, they are either not targeting NPU or have limited scope to explore AI workload.
\cite{Yang2017AMT, 8942149} focus specifically on NPUs with power modeling capability for both low level components and large design hierarchies.
But the activity factors of AI workloads are extracted offline rather than during runtime with the corresponding performance analysis.

The VPU-EM framework provides a simulation mode called Power-EM to enable joint performance and power analysis seamlessly with scalable AI workloads.
Power-EM leverages the power characterization data extracted from the actual backend design implementation of hardware components of VPU for a given target process node.
In Power-EM mode, the framework calculates fine-grained power characteristics of the hardware submodules based on the AI workload processing activities extracted for the sub-modules
within small user-defined time intervals. This capability allows both peak and average power to be simulated down to submodule level, factoring concurrency and scheduling
efficiency of the actual AI workloads. In this section, we present the detailed description of Power-EM mode.

\subsection{Power-EM Methodology}
\label{subsec:power-methodology}
Power-EM mode takes a hierarchical design description from a yaml configuration file. Each design hierarchy is represented by a power node which contains the power characterization
data of the corresponding design. Power nodes can contain sub-nodes and top level logic. During simulation, each power node instance is bonded to the performance model of
the corresponding hardware module to collect module-specific power simulation statistics.

Power-EM relies on power characterization data extracted from the backend implementation of a hardware module using conventional EDA-based power simulation tools such as PrimePower.
Power for each power node contains two components, leakage power and dynamic power:
$$
P_{total} = P_{lkg} + P_{dyn}
$$

Leakage power characterization includes a leakage power value of the power node characterized under nominal voltage and temperature operating conditions, and a process-dependent
lookup table. The lookup table is used to adjust the leakage power dynamically based on a different operating condition set in simulation:

$$
P_{lkg} = P_{lkg0} \times \frac {LkgRatio\_LUT(temp, voltage)} {LkgRatio\_LUT(temp_{0}, voltage_{0})}
$$

Dynamic power is computed based on the formula:
$$
P_{dyn} = C \times F \times V^2
$$
where C, F and V are switching capacitance, operating frequency and voltage respectively.

In Power-EM, $P_{dyn}$ is further divided into two parts, a static part that is independent of workload processing activities (to account of
clock switching etc.) and a variable part that is dependent on workload processing activities.

Similar to leakage power characterization, we run backend power simulation flow under a nominal operating condition of frequency, temperature and voltage. We extract the
equivalent switching capacitance of a given module for both the static part and the variable part as $C_{dyn\_idle}$ and $C_{dyn\_active}$ respectively. For $C_{dyn\_active}$,
synthetic workload is used to generate the maximum switching factor of the design. A process-dependent VF curve is characterized to provide operating condition scaling during simulation.

In Power-EM mode simulation, the dynamic power is computed with an effective voltage scaled from the VF curve.
The workload-dependent part of the dynamic power is further scaled using the utilization statistics collected from the actual workload.

$$
V_{adj} = f2v(F, T)
$$
$$
P_{dyn} = (Cdyn\_idle + Cdyn\_active \times utilization) \times F \times V_{adj}^2
$$

In VPU, a significant portion of the dynamic power is depending on the switching activities of different hardware modules triggered by the specific workloads. Due to processing
concurrency and resource dependency, the switching activities are very dynamic. Power-EM allows user to specify a time interval, called power trace interval (PTI), for the activity statistics to
collected based on VPU-EM performance simulation. The event-driven nature of the simulation allows the statistics to be captured both spatially and temporally. Utilization for a specific
module instance and a specific PTI is computed based on the corresponding activity data and the maximum activity of the hardware capability. Unlike RTL or gate level power simulation tools,
activity and utilization in Power-EM mode are computed based on hardware events, which allows Power-EM to scale the simulation speed to analyze many AI models against many architecture configurations.
Table \ref{tab:tab2} defines how each hardware module generates activity statistics.

\begin{table}[H]
  \caption{Maximum and Measured Activities}
  \centering
  \begin{tabular}{llll}
    \toprule
    \textbf{Hardware} & \textbf{Maximun}   & \textbf{Measured}  \\
    \textbf{Module}   & \textbf{Activity}  & \textbf{Activity} \\
    \midrule
    \multirow{ 2}{*}{DMA} & maximum & measured \\
                          & data transfer BW & data transfer BW \\ \hline
    \multirow{ 2}{*}{NOC} & maximum & measured \\
                          & data transfer BW & data transfer BW \\ \hline
    \multirow{ 2}{*}{CB} & maximum & measured \\
                          & data access size & data access size \\ \hline
    \multirow{ 2}{*}{DDR} & maximum & measured \\
                          & data access size & data access size \\ \hline
    \multirow{ 2}{*}{DPU} & ideal & processed \\
                          & op count & op count \\ \hline
    \multirow{ 2}{*}{DPU} & ideal & processed \\
                          & op count & op count \\
    \bottomrule
  \end{tabular}
  \label{tab:tab2}
\end{table}

\subsection{Joint Performance/Power Analysis}
\label{subsec:power-analysis}

\begin{figure}[H]
  \centering
  \includegraphics[scale=0.38]{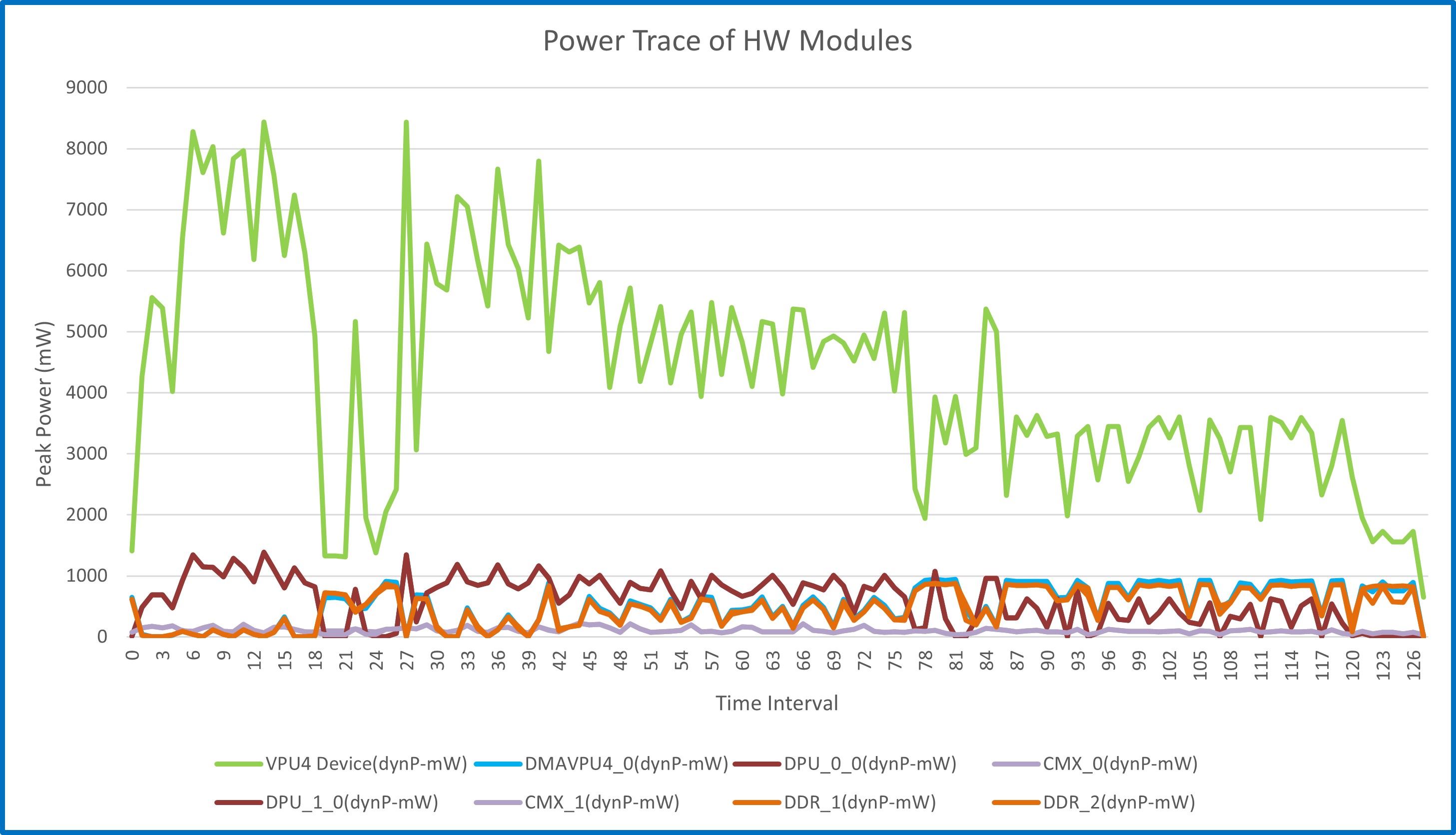}
  \caption{Power Profiling in Power-EM Mode}
  \label{fig:fig9}
\end{figure}

Power-EM mode simulation can provide detailed power profiles of the hardware modules of VPU for a given real AI workload as shown in Figure \ref{fig:fig9}.
In this analysis, the transient power of the selected hardware modules are calculated based on a user-defined PTI. It provides detailed correlation between
different AI tasks and the corresponding processing engines. Targeted architecture improvements can be made to specific engines to improve performance without
increasing processing power or reduce power consumption while maintaining performance.

\begin{figure}[H]
  \centering
  \includegraphics[scale=0.42]{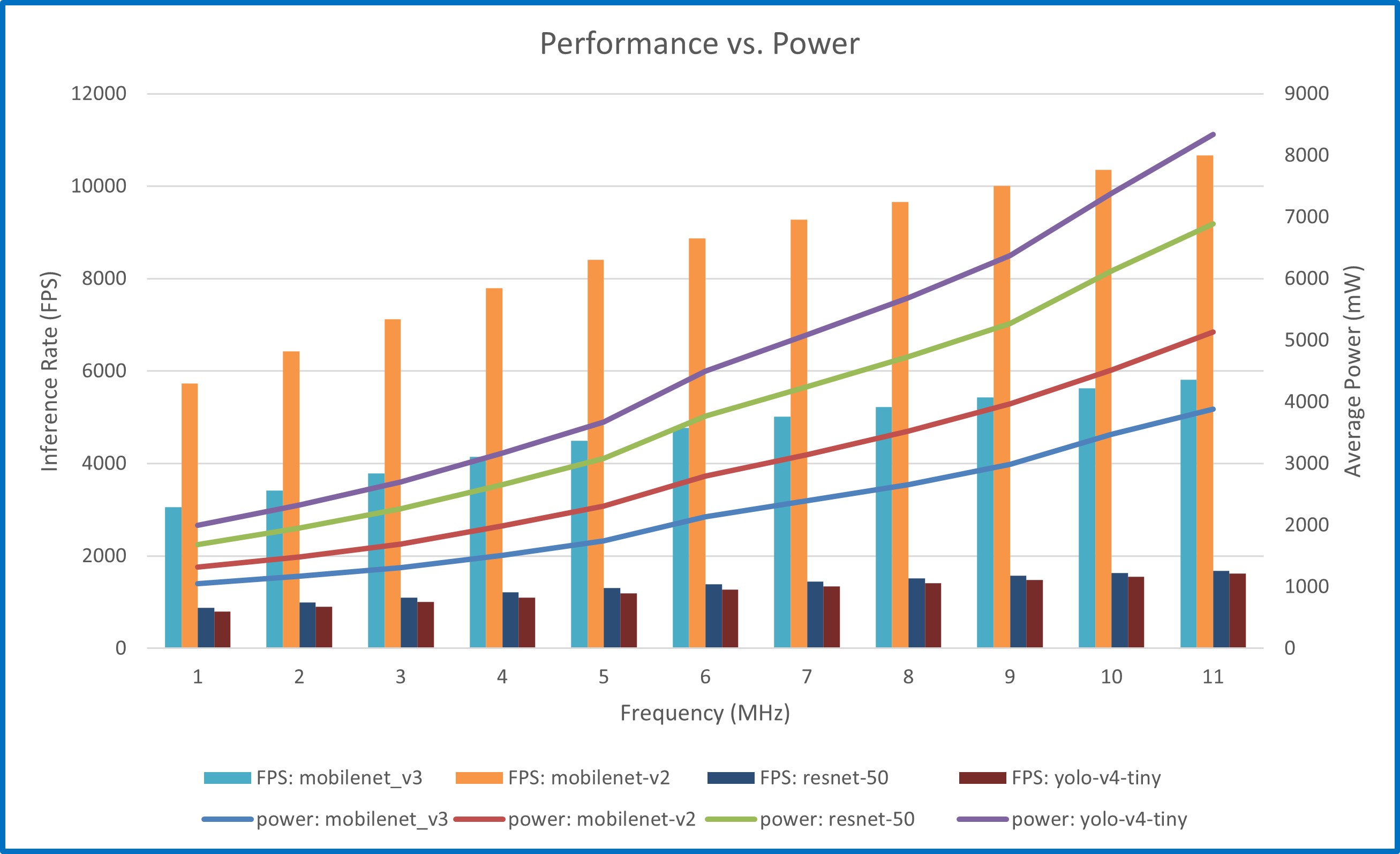}
  \caption{Joint Performance/Power Analysis}
  \label{fig:fig10}
\end{figure}

A second example is shown in Figure \ref{fig:fig10}, where Power-EM mode is scaled to perform a joint performance/power analysis for a set of AI models
across a wide range of operating conditions. In this example, we sweep the operating frequency of DPU through a wide range with 100MHz steps. The operating voltage
is computed by Power-EM mode using the pre-characterized VF curve. The inference performance and average power consumption of the models are obtained simultaneously
in the same analysis for all operating frequencies. Leveraging this result, workload-specific DVFS algorithms can be developed to optimize device battery life without sacrificing
 minimum performance requirements.

\section{Conclusion and Future Work}
\label{sec:conclusion and future}

\subsection{Conclusion}
\label{subsec:conclusion}

In this paper, a novel NPU performance/power modeling framework, VPU-EM is presented using Intel VPU architecture as a reference target. We analyzed
the complexity and challenges facing NPU architects to conduct performance/power analysis effectively. This is further highlighted by
the limitations of the existing modeling approaches. We also asserted that the analysis for NPU must be conducted using full model inference
with diversified AI workloads to explore the large design parameter space sufficiently.
A detailed description of the performance/power modeling methodology deployed by VPU-EM is subsequently presented to address the challenges.
Through several concrete examples, we demonstrated the comprehensiveness and scalability of the VPU-EM framework.
We conclude that a comprehensive performance/power modeling methodology like VPU-EM is effective to tackle the complexity of the NPU architecture.
Furthermore, even though VPU-EM is developed specifically for Intel VPU, its methodology can be generalized and applied to architecture research of NPUs at large.

\subsection{Future Work}
\label{subsec:future}
The ongoing work to enhance the VPU-EM framework includes two critical aspects:

\begin{itemize}
  \item the addition of Stack-EM mode to analyze the performance impacts of different layers of the software stack with multi-context use case based scheduling pipeline
  \item the enhancement of Power-EM mode to analyze the performance/power characteristics under active power state management and DVFS for given use cases
\end{itemize}

\normalsize
\bibliography{references}


\end{document}